# Machine Learning Based Animal Emotion Classification Using Audio Signals

Mariia Slobodian
*Wallee AG, Winterthur, Switzerland*

Mykola Kozlenko
*SoftServe, Austin, USA*

*Abstract*—**This paper presents the machine learning approach to the automated classification of a dog's emotional state based on the processing and recognition of audio signals. It offers helpful information for improving human-machine interfaces and developing more precise tools for classifying emotions from acoustic data. The presented model demonstrates an overall accuracy value above 70% for audio signals recorded for one dog.**

## I. Introduction

Scientists suggest that canines are far more intelligent than people realize. Over the years, there have been a lot of publications about studies focused on dogs. Studies have revealed that dogs demonstrate advanced emotional intelligence. They are equipped to perform great deeds for their owners and have a high emotional sensitivity. Accordingly, it is reasonable to assume that humans have long felt the desire to speak with dogs. In addition, there have been a lot of attempts to handle and process the audio signals. It's interesting how vocal emotion recognition software functions because it resembles facial expression detection software in many ways. Applying human voice emotion recognition techniques to dog recordings also shows promising results. The best accuracy so far is ultimately insufficient for use in practical applications.

## II. Related works

A number of similar research works in the field of modern human-machine interfaces have used recognition systems, such as human voice recognition by emotion and gender, and they have shown strong accuracy results. Emotional prosody can be used to identify emotions, according to research on human speech. [1]. This study used comparisons with other live species to examine the evolution of speech from a comparative and phylogenetic perspective. People can recognize barks that are related to circumstances such as being willing to pay, being angry, or being aggressive [2]. It was proved that dog barks are the most well-known animal voice to humans [3]. Furthermore, barks have specific acoustic features [4]. Changes in an animal's emotional state should have an impact on the muscle systems that govern its vocal apparatus, changing the acoustic characteristics of vocal transmission, especially when the animal engages in safety defense [4].

A preliminary investigation of machine learning has revealed that auditory descriptors are appropriate for detecting bark context, with a classification efficiency of 43% [5]. Different acoustic features that are known to reflect human emotions were compared in order to divide canine bark patterns according to their context or observed emotion. Publications [6], [7], [8] offer a method that effectively separates relevant information from useful signals and random noise using deep learning and artificial neural networks. Applying those methods to categorize the emotions of dogs seems promising.

III. Methodology

During our research, we identified two key methods with typical data preprocessing approaches such as Mel-frequency cepstral coefficients (MFCC) and Extractor Discovery System (EDS). It demonstrates that while MFCC is primarily used to obtain information from audio files and depict it using pictures, EDS iteratively generates descriptor generations and can mark the best-performing descriptors from one generation as seeds for subsequent ones. To solve this problem, EDS employs a range of artificial intelligence tools. First off, a type system forbids the creation of descriptors that include mistakes. In order to further elaborate, new descriptors are employed as the second phase. On the other hand, by using some effective classifiers, MFCC gives us 13 relevant coefficients at the end that we can use with appropriate machine learning techniques. But in this research, we do not use any preprocessing and feature engineering steps. Audio signals pass directly to the convolutional neural network (CNN).

We found that the ideal set of emotions contains five types of feelings (Fig. 1), which are fewer compared to human measurements. Those emotion classes served well our needs in model classification. Furthermore, it requires extra effort to analyze the dynamic movement of the vocal folds. The following voice cues could be used to identify the vocal characteristics: timbre, amplitude, loudness, intensity, and spectral composition.

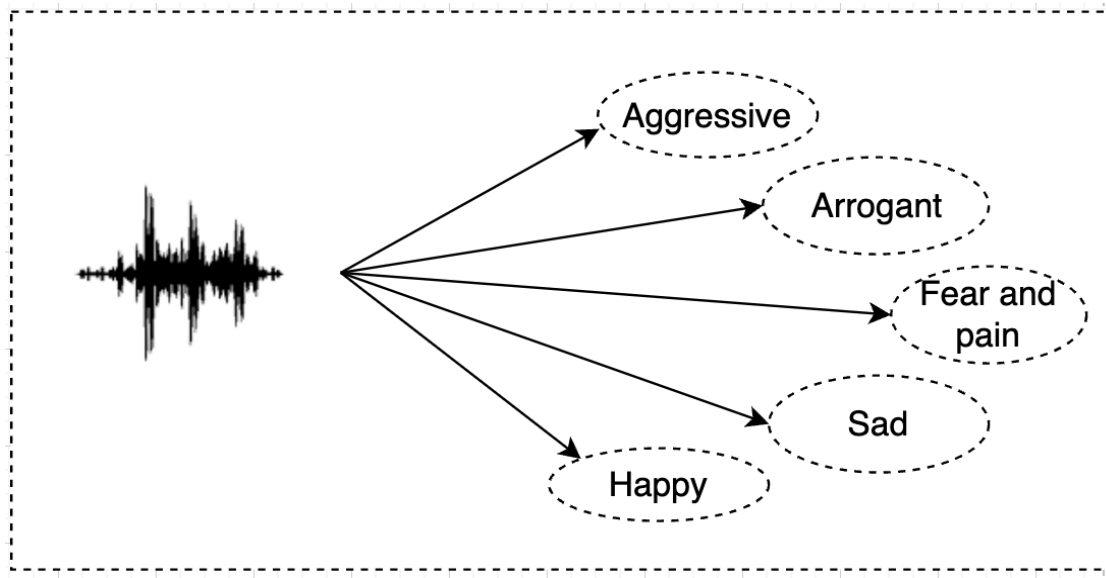

Fig.1. Emotion classes

The structure of the developed deep learning model is shown in Fig. 2. It gets the fragment of the audio signal at the input. The length of the signal is a hyperparameter (12 000 samples, for example). It contains two hidden convolutional layers with a ReLU activation function and one fully-connected layer with five output neurons and softmax activation. Batch normalization is used between all layers. The model outputs five confidence values for each emotion class (Fig. 2.). The model was trained with 4000 signal fragments. The size of the validation set is 800 records.

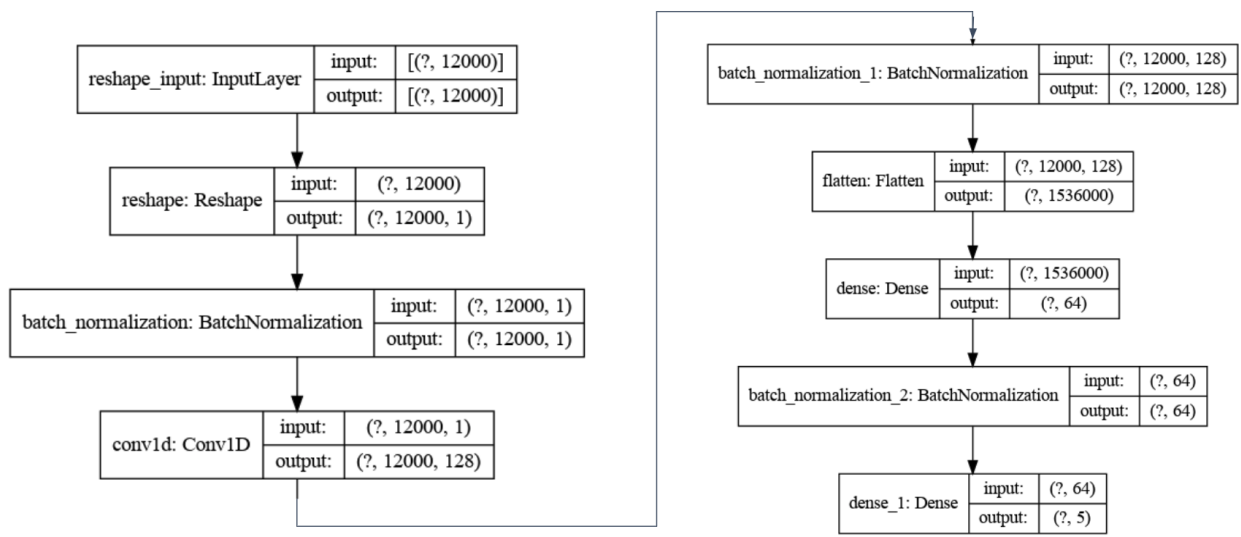

Fig. 2. Structure of the deep learning model

The model performance assessment is performed as a post-predict evaluation. It demonstrates how closely the predictions represent the actual ground truth. We use a widely-used classification report that outputs class-wise precision, recall, f1-score as well as overall accuracy. The size of the test set is 1500 records.

## IV. Results

The classification report is shown in Fig.3. The overall accuracy is about 72% (model trained from one dog).

|  | precision | recall | f1-score | support |
|---|---|---|---|---|
| aggressive | 0.73 | 0.81 | 0.77 | 280 |
| arrogant | 0.70 | 0.61 | 0.65 | 288 |
| fear_and_pain | 0.66 | 0.93 | 0.77 | 302 |
| happy | 0.80 | 0.54 | 0.64 | 329 |
| sad | 0.76 | 0.74 | 0.75 | 301 |
|  |  |  |  |  |
| accuracy |  |  | 0.72 | 1500 |
| macro avg | 0.73 | 0.73 | 0.72 | 1500 |
| weighted avg | 0.73 | 0.72 | 0.72 | 1500 |

Fig. 3. Classification report

## V. Discussion

The variety of feelings could be recognized by us, but in comparison to positive ones, the proportion of negative vocalizations was dominant. It can be challenging to elicit neutral vocalizations because anger and playfulness are typically active emotions. Despite that, the joyful class of barking is still the most recognizable despite the limited data. It is challenging to train models because of the lack of labeled data of good quality. Different outcomes were seen when the same model was applied to the entire group of dogs as opposed to each individual dog.

## VI. Future research

The experiments have demonstrated the potential of using machine learning models to distinguish between various bark properties. However, it seemed that the attained accuracy wasn't good enough for applications in the real world. Additional training and improvement are needed to achieve greater results. Soon, we'll investigate how well-known feature extraction techniques can be linked with deep learning strategies as well as the direct usage of row signals as features. It could greatly reduce the number of computer resources needed and streamline the data processing pipeline. Analysis of the potential use of recurrent neural networks and architectures is also necessary.

## VII. Conclusion

We examined the possibility of distinguishing dog bark using acoustic feature sets and affective computing based on context, perceived emotion, and intensity. Moreover, we offered recommendations for how to enhance the examined techniques for classifying emotions. It must be emphasized that the approach analysis offered here could make a substantial impact and excite business interest.

## VIII. Acknowledgment

The authors sincerely acknowledge the assistance of the experts of the Department of Information Technology of the Vasyl Stefanyk Precarpathian National University for their technical support during the real analysis as well as their professional guidance for consultations.

## IX. Disclosures

The authors declare that there are no conflicts of interest related to this paper.

## References

[1] W.T. Fitch, "The biology and evolution of speech: A comparative analysis," *Annual Review of Linguistics*, vol. 4, pp. 255–279. 2018, doi: 10.1146/ANNUREV-LINGUISTICS-011817-045748.

[2] S. Hantke, N. Cummins, and B. Schuller, "What is my Dog Trying to Tell Me? the Automatic Recognition of the Context and Perceived Emotion of Dog Barks," *2018 IEEE International Conference on Acoustics, Speech and Signal Processing (ICASSP)*, 2018, pp. 5134-5138, doi: 10.1109/ICASSP.2018.8461757.

[3] D. U. Feddersen-Petersen, "Vocalization of European wolves (*canis lupus lupus* L.) and various dog breeds (*canis lupus* F. Fam.)," *Archives Animal Breeding*, vol. 43, no. 4, pp. 387–398, 2000, doi: 10.5194/AAB-43-387-2000.

[4] S. Yin, "A new perspective on barking in dogs (canis familaris.)" *Journal of Comparative Psychology*, vol. 116, no. 2, pp. 189–193, 2002, doi: 10.1037/0735-7036.116.2.189.

[5] C. Molnar, F. Kaplan, P. Roy, F. Pachet, P. Pongracz, A. Doka, and A. Miklosi, "Classification of Dog Barks: A Machine Learning Approach," *Animal Cognition*, vol. 11, pp. 389–400, 2008, doi: 10.1007/S10071-007-0129-9

[6] M. Kozlenko, I. Lazarovych, V. Tkachuk, and V. Vialkova, "Software Demodulation of Weak Radio Signals using Convolutional Neural Network," *2020 IEEE 7th International Conference on Energy Smart Systems (ESS)*, 2020, pp. 339-342, doi: 10.1109/ESS50319.2020.9160035

[7] M. Kozlenko and V. Vialkova, "Software Defined Demodulation of Multiple Frequency Shift Keying with Dense Neural Network for Weak Signal Communications," *2020 IEEE 15th International Conference on Advanced Trends in Radioelectronics, Telecommunications and Computer Engineering (TCSET)*, 2020, pp. 590-595, doi: 10.1109/TCSET49122.2020.235501

[8] I. Lazarovych et al., "Software Implemented Enhanced Efficiency BPSK Demodulator Based on Perceptron Model with Randomization," *2021 IEEE 3rd Ukraine Conference on Electrical and Computer Engineering (UKRCON)*, 2021, pp. 221-225, doi: 10.1109/UKRCON53503.2021.9575458.